\documentclass{article}
\usepackage{spconf,amsmath,graphicx}
\usepackage{url}
\usepackage{adjustbox}
\usepackage{array}
\usepackage{booktabs}
\usepackage{multirow}
\usepackage{amsmath}
\usepackage{array}
\usepackage{booktabs}
\usepackage{float}
\usepackage{adjustbox}
\usepackage{stfloats} 
\usepackage{amsmath}
\usepackage{cite}
\usepackage{romannum}
\usepackage{xcolor}

\title{Speaker Contrastive Learning for Source Speaker Tracing}
\name{Qing Wang$^{1,2}$, Hongmei Guo$^{2,3}$, Jian Kang$^2$, Mengjie Du$^2$, Jie Li$^2$, Xiao-Lei Zhang$^{2,3}$, Lei Xie$^{1*}$\thanks{* Corresponding author.}}
\address{ $^1$Audio, Speech and Language Processing Group (ASLP@NPU), School of Computer Science, \\ Northwestern Polytechnical University, Xi’an, China\\
$^2$Institute of Artificial Intelligence (TeleAI), China Telecom, China \\
$^3$School of Marine Science and Technology, Northwestern Polytechnical University, Xi'an, China
  }
\begin{document}
%
\maketitle
\begin{abstract}

As a form of biometric authentication technology, the security of speaker verification (SV) systems is of utmost importance. However, SV systems are inherently vulnerable to various types of attacks that can compromise their accuracy and reliability. One such attack is voice conversion (VC), which modifies a person’s speech to sound like another person by altering various vocal characteristics. This poses a significant threat to SV systems.
To address this challenge, the Source Speaker Tracing Challenge (SSTC) in IEEE SLT2024 aims to identify the source speaker information in manipulated speech signals. Specifically, SSTC focuses on source speaker verification against voice conversion to determine whether two converted speech samples originate from the same source speaker. 
In this study, we propose a speaker contrastive learning-based approach for source speaker tracing to learn the latent source speaker information in converted speech. To learn a more source-speaker-related representation, we employ speaker contrastive loss during the training of the embedding extractor. This speaker contrastive loss helps identify the true source speaker embedding among several distractor speaker embeddings, enabling the embedding extractor to learn the potentially possessing source speaker information present in the converted speech.
Experiments demonstrate that our proposed speaker contrastive learning system achieves the lowest EER of 16.788\% on the challenge test set, securing first place in the challenge.
\end{abstract}
\section{Introduction}
\label{sec:intro}

Speaker verification (SV) is a critical biometric authentication technology used in various daily applications, such as security systems, forensic investigations, and user authentication for various services. Its significance lies in ensuring secure and personalized user interactions, making the security of SV systems of utmost importance. 

Although SV technology has been widely adopted in various fields, these SV systems~\cite{snyder2018x, desplanques20} are inevitably vulnerable to various types of attacks that can compromise their accuracy and reliability.
Adversarial attacks~\cite{goodfellow2014explaining, carlini2017towards, szegedy2013intriguing, kurakin2016adversarial}, for instance, aim to confuse the SV system by introducing well-crafted perturbations into the speech signals. In recent years, many studies~\cite{wang2019adversarial, kreuk2018fooling, wang2020inaudible, abdullah2019practical, wang2023timbre, wang2023pseudo} have conducted effective adversarial attacks on SV systems. 
On the other hand, spoofing attacks~\cite{wu2012detecting, wu2015spoofing} attempt to mimic the timbre of target speakers through methods commonly including impersonation, replay, voice conversion, and speech synthesis. 
Spoofing attacks can be broadly classified into two categories: physical access attacks and logical access attacks. In physical access attacks, the samples are input into the SV system through a sensor or microphone (source) level, directly attacking the system. In contrast, logical access attacks involve direct injection into the SV system, with the most common methods being speech synthesis and voice conversion.

\begin{figure}
    \centering
    \includegraphics[width=1\linewidth]{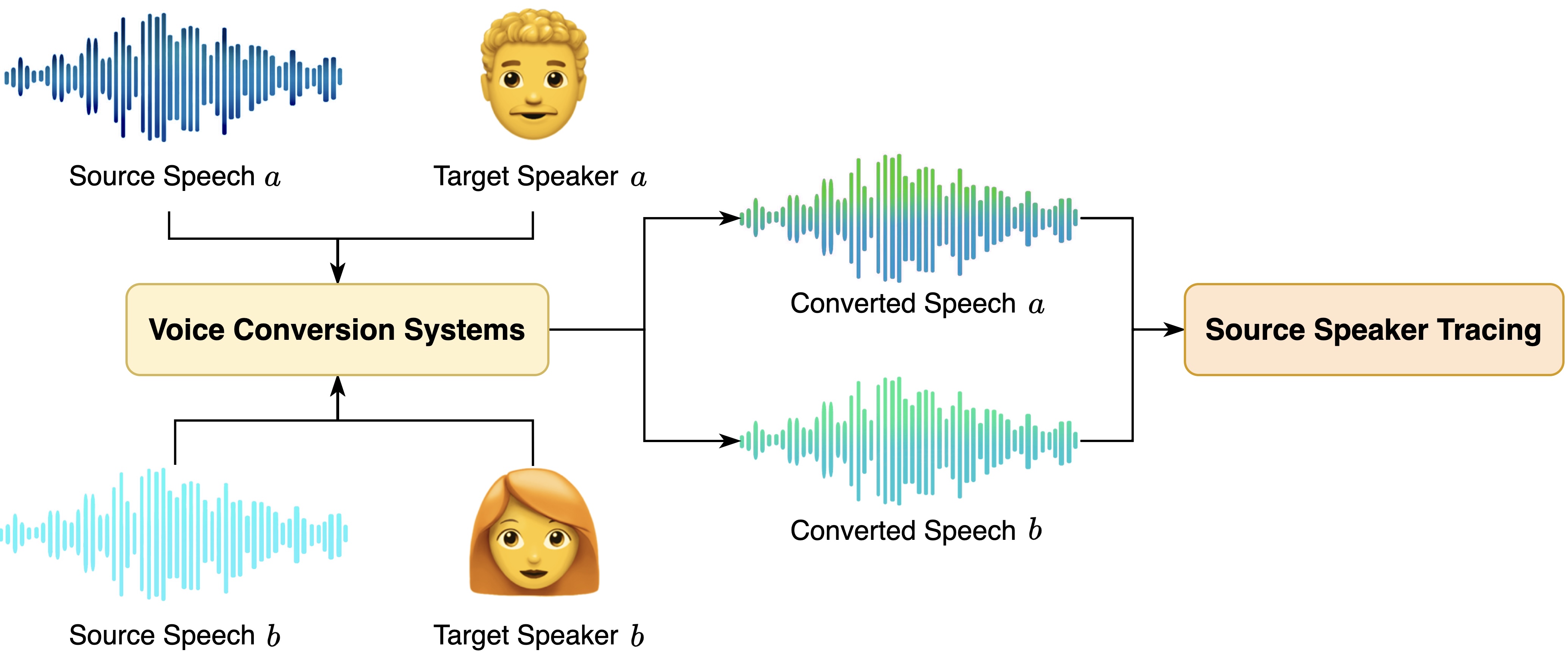}
    \caption{Source speaker verification against voice conversion.}
    \vspace{-8pt}
    \label{fig:sourcesv}
\end{figure}

As a typical type of spoofing attack, voice conversion (VC)~\cite{oord2016wavenet,ren2019fastspeech} is a technology that modifies the characteristics of a person's speech (source speaker) to make it sound like another person's (target speaker) voice while retaining the original content, posing a significant threat to SV systems.
To address the threat of VC in speaker verification, the SLT2024 Source Speaker Tracing Challenge (SSTC)\footnote{https://sstc-challenge.github.io/} has introduced the task of source speaker verification against voice conversion. As shown in Figure~\ref{fig:sourcesv}, given a speech utterance from a source speaker and a speech utterance from a target speaker, voice conversion (VC) manipulates the speech signal of the source speaker to make it sound like the target speaker while preserving the original linguistic content. Participants are required to develop models that can extract information about the source speaker from the converted speech and determine whether two converted utterances originate from the same source speaker. Source speaker tracing aims to determine whether the source speakers of converted speech are the same, as converted speech retains certain aspects of the source speaker’s speaking style~\cite{cai2023identifying}.

In~\cite{cai2023identifying}, Cai \textit{et al.} indicated that current voice conversion techniques are not perfect enough, as the converted speech retains certain characteristics of the source speaker’s speech style. Building on this observation, they successfully introduced a method for source speaker verification by training models on converted audio labeled with the source speaker information.
As discussed in~\cite{li2024database}, beyond the source speaker information, the converted speech also contains traces of the conversion methods used. Given the distinct differences among various VC methods, it is possible to extract these conversion traces to identify the specific method employed and to further process the converted speech. Li \textit{et al.} utilized a multi-task learning approach to achieve both source speaker verification and conversion method recognition. Inspired by this finding, we take certain characteristics of the source speaker’s speech style in converted speech retained as latent source speaker information and consider whether the potential source speaker information can be better learned in the embedding extractor.


To address the challenge of source speaker tracing in the context of voice conversion, we propose a speaker contrastive learning approach. This method employs speaker contrastive loss to enhance model training, allowing the generation of more discriminative embeddings related to source speakers. Our approach focuses on the latent source speaker information retained in converted speech, aiming to better capture this information using an embedding extractor.
To reveal the latent source speaker information in the embeddings of converted speech, we integrate a speaker embedding extractor trained with converted speech data. This extractor learns to capture the source speaker's characteristics that persist after conversion. The speaker contrastive loss is used to identify the true source speaker embedding among several distractor embeddings, ensuring that the embedding extractor effectively learns the source speaker information present in the converted speech.
Experimental results on the SLT2024 SSTC datasets demonstrate that our proposed system significantly outperforms all evaluated systems, securing first place in the challenge.

\section{Related Work}
\label{sec:related}

In this section, we provide an overview of the existing solutions~\cite{cai2023identifying,li2024database} to source speaker verification against voice conversion.
Both approaches leverage the commonalities among converted speech from various voice conversion algorithms, enabling general source speaker identification.

\subsection{Speaker Embedding Clustering}
\label{sec:2.1}

MFA-Conformer~\cite{zhang2022mfa, cai2023leveraging} enhances the speaker embedding network by adapting the speaker embedding space to address the challenges posed by voice conversion techniques. This network aims to create a distinct embedding space where utterances from the same speaker are grouped, while those from different speakers are separated. Voice conversion models alter the source speech to resemble the target speaker’s voice, often placing converted speech within the target speaker’s subspace. Source speaker identification seeks to map the converted speech back to the source speaker’s subspace.

Starting with a source speech dataset $\mathcal{D}_s = \{s_i\}$ and a target speech dataset $\mathcal{D}_t = \{t_j\}$, a voice conversion model modifies the voice of each source speech $s_i$ to sound like the voice of a target speech $t_j$, resulting in a converted speech dataset $\mathcal{D}_c = \{x_{s_i \rightarrow t_j} \mid s_i \in \mathcal{D}_s, t_j \in \mathcal{D}_t\}$. To train the speaker embedding network, the datasets $\mathcal{D}_s$, $\mathcal{D}_t$, and the converted speech datasets $\mathcal{D}_{c,k}$ generated by the voice conversion algorithms $C_k$ (for $k = 1, \ldots, K$) are combined into the training dataset $\mathcal{D} = \mathcal{D}_s \cup \mathcal{D}_t \cup \mathcal{D}_{c,1} \cup \cdots \cup \mathcal{D}_{c,K}$. During the training process, the label assigned to each converted speech $x_{s_i \rightarrow t_j}$ is the speaker identity of the source speech $s_i$. The speaker embedding network is thus optimized to capture the source speaker's characteristics from the converted speech while maintaining a discriminative embedding space.

\subsection{Adapter-based Multi-Task Learning}
\label{sec:2.2}
As discussed in~\cite{li2024database}, different VC methods leave different traces in the generated fake audio. These conversion traces are analyzed to identify the specific method employed and to further process the converted speech.
The MFA-Conformer with an adaptor improves the ability to distinguish converted speech by different kinds of VC methods, as these methods leave different traces in the generated fake audio. The adaptor enhances performance by fine-tuning the model to better adapt to specific datasets or tasks, enabling more effective identification of these traces.

The integration of source speaker verification and conversion method recognition tasks acknowledges their inherent connection. Despite their distinct objectives, these tasks share interrelated information present in converted speech samples. Multi-task learning provides a unified approach to tackle both objectives concurrently. By utilizing task-specific parameter sets embedded within adapter modules in the model architecture, shared learning across tasks is facilitated, enabling the model to simultaneously identify both the source speaker and the specific conversion method.

During training, the model learns to distinguish between known and unseen conversion methods. Utterances from the same conversion method tend to cluster together, whereas those from different methods exhibit greater dispersion. This clustering effect aids in recognizing known methods included in the training set. However, identifying unseen methods poses a challenge due to the impracticality of encompassing all potential techniques in the training data. Nevertheless, the model effectively utilizes similarities in Euclidean distances between audio samples to distinguish between known and unseen conversion methods. This approach equips the model to robustly handle both types of methods during inference, thereby enhancing its practical utility in real-world applications.

\section{Speaker Contrastive Learning}
\label{sec:method}
In this section, we detail the proposed speaker contrastive learning method for the source speaker tracing task. This method employs speaker contrastive loss to enhance the embedding extractor training, enabling the generation of more discriminative embeddings related to source speakers. 

We integrate a speaker embedding extractor, trained with converted speech data, to capture the source speaker’s characteristics that remain after conversion. The embedding extractor for the source speaker tracing task is trained in three continuous phases:
\begin{itemize}
    \item Phase \Romannum{1}: The model is initially trained using the Librispeech train-clean set (only source speech). 
    \item Phase \Romannum{2}: The training set of SSTC 2024 (converted speech) and Librispeech data (source speech) are both used to fine-tune the model. 
    \item Phase \Romannum{3}: The model is trained using only converted speech, employing both speaker AAM-Softmax loss and contrastive loss with the source speech embeddings extracted by the fixed Phase \Romannum{1} model.
\end{itemize}

This three-phase training procedure for the embedding extractor incorporates both source and converted speech, which improves the model’s ability to learn the latent source speaker information in the converted speech enhancing source speaker tracing by gradually fine-tuning and specializing the model.

\begin{figure}[h]
  \centering
  \includegraphics[width=8.8cm]{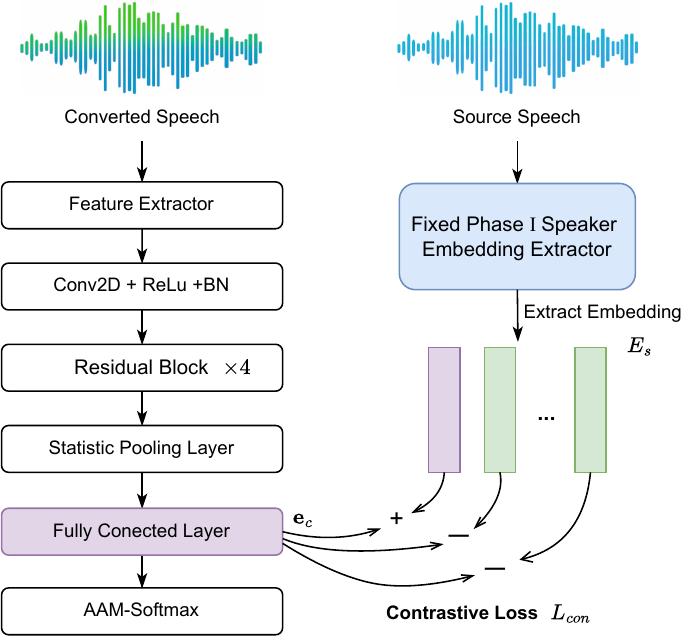}
  \caption{Overview of the proposed source speaker contrastive learning system. $\mathbf{e}_c$ and $E_s$ refer to the representation of the embedding extraction layer and a set of embeddings of source speech, respectively. Here, $+$ and $-$ represent positive and negative sample pairs, respectively, which are used to compute the speaker contrastive loss $\mathcal{L}_{\text{Con}}$.}
  \label{fig:conloss}
\end{figure}

The speaker contrastive learning training procedure of Phase \Romannum{3} is outlined in Figure~\ref{fig:conloss}.
After the Phase \Romannum{1} speaker embedding extractor is trained, it remains fixed to extract embeddings from source speech (as shown in the right part of Figure\ref{fig:conloss}). 
These embeddings are used as positive and negative samples to compute the speaker contrastive loss. Subsequently, the Phase \Romannum{2} speaker embedding extractor network is further trained using converted speech, with the representation from a fully connected layer employed for computing the speaker contrastive loss.

The speaker contrastive loss is adopted to identify the true source speaker embedding within $K$ distractors of speaker embeddings so that the embedding extractor can learn the potentially possessing source speaker information in the converted speech.
Given a converted speaker embedding $\mathbf{e}_c$ and a set of source speaker embeddings $E_s = \{\mathbf{e}_s^1, \dots, \mathbf{e}_s^K\}$, the contrastive loss can be calculated as:
\begin{equation} \label{eq:con_loss}
    \mathcal{L}_{\text{Con}}  = - \log \frac{\exp (\cos(\mathbf{e}_c, \mathbf{e}_k) / \tau)}  {\sum_{e_s \sim E_s} \exp (\cos( \mathbf{e}_c, \mathbf{e}_s) / \tau )},
\end{equation}
where $\cos \left( \cdot \right)$ represents the cosine similarity between two vectors and $\tau$ is a temperature hyper-parameter. 
Here, $E_s$ is a set comprising all candidate source speaker embeddings, where the positive sample $\mathbf{e}_k$ belongs to the same speaker as converted embedding $\mathbf{e}_c$ and others are negative samples from different speakers.

The final loss function of our source speaker contrastive learning model is given by:
\begin{equation}
    \mathcal{L} = \mathcal{L}_{\text{AAM}} + \alpha \mathcal{L}_{\text{Con}},
\end{equation}
where $\mathcal{L}_{\text{AAM}}$ is the loss function of the speaker embedding extractor, $\alpha$ refers to an interpolation factor that scales the speaker contrastive loss.
We sample $K=5$ negative samples during the training and set $\alpha = 1$ to ensure that the two losses are of a similar magnitude. By identifying the true source speaker embedding among several distractor speaker embeddings, the embedding extractor can better learn the source speaker information potentially present in the converted speech.

\section{EXPERIMENTAL SETUP AND RESULTS}
\label{sec:experience}

In this section, the datasets and experimental setup and implementation details of source speaker tracing models are introduced.  
Besides, we also present the experimental results and analysis of our proposed method.

\subsection{Dataset}

SLT2024 SSTC focuses on source speaker verification against voice conversion, aiming to determine whether two converted speech samples originate from the same source speaker. 
As shown in Table~\ref{tab:vc16}, the SSTC dataset is constructed from fake audio generated by 16 different VC models. Table~\ref{tab:vc16} lists the datasets used for training, development, and testing for each of the VC methods, along with their respective GitHub repositories (all the repositories can be accessed at the \url{https://*.github.com}). Among them, 8 methods are used in the training set of the challenge. For the Dev set, 4 additional methods are included compared to the train set. The test set includes four more unknown methods (unknown during the competition) compared to the Dev set.

\begin{table}[ht]
\centering
\caption{Train and test sets and repositories for each VC method \cite{li2024database}.}
\vspace{5pt}
\small
\begin{adjustbox}{max width=0.48\textwidth}
\begin{tabular}{@{}llccc@{}}
\toprule
\textbf{Method} & \textbf{Train set} & \textbf{Dev set} & \textbf{Test set} & \textbf{Repository} \\ \midrule
AGAIN-VC \cite{chen2021again} & Train-1 & Dev-1 & Test-1 & KimythAnly/AGAIN-VC \\
FreeVC \cite{li2023freevc} & Train-2 & Dev-2 & Test-2 & OlaWod/FreeVC \\
MediumVC \cite{gu2021mediumvc} & Train-3 & Dev-3 & Test-3 & BrightGu/MediumVC \\
StyleTTS \cite{li2023styletts} & Train-4 & Dev-4 & Test-4 & yl4579/StyleTTS-VC \\
TriAAN-VC \cite{park2023triaan}& Train-5 & Dev-5 & Test-5 & winddori2002/TriAAN-VC \\
VQMIVC \cite{wang2021vqmivc} & Train-6 & Dev-6 & Test-6 & Wendison/VQMIVC \\
SigVC\cite{zhang2022sig} & Train-7 & Dev-7 & Test-7 & - \\
KNN-VC \cite{baas2023voice}& Train-8 & Dev-8 & Test-8 & bshall/knn-vc \\
BNE-PPG-VC \cite{liu2021any} & - & Dev-9 & Test-9 & liusongxiang/ppg-vc \\
DiffVC  \cite{popov2021grad}& - & Dev-10 & Test-10 & huawei-noah/Speech-Backbones \\
S2VC \cite{lin2021s2vc}& - & Dev-11 & Test-11 & howard1337/S2VC \\
YourTTS \cite{casanova2022yourtts}& - & Dev-12 & Test-12 & Edresson/YourTTS \\
ControlVC \cite{chen2022controlvc}& - & - & Test-13 & MelissaChen15/control-vc \\
Diff-HierVC \cite{choi2023diff}& - & - & Test-14 & hayeong0/Diff-HierVC \\
LVC-VC \cite{kang23b_interspeech}& - & - & Test-15 & wonjune-kang/lvc-vc \\
Wav2vec-VC \cite{lim2024wav2vec}& - & - & Test-16 & prairie-schooner/wav2vec-vc \\ \bottomrule
\end{tabular}
\end{adjustbox}
\label{tab:vc16}
\end{table}

The Librispeech dataset~\cite{panayotov2015librispeech} is used as source speakers, while the VoxCeleb datasets~\cite{nagrani2017voxceleb, chung2018voxceleb2} are used as target speakers to generate converted speech using the aforementioned VC models. 
Specifically, the train-clean, dev-clean, and test-clean sections of the Librispeech dataset are used as the source speech to construct the training set, development set, and test set of the converted speech dataset, respectively. For target speech, the VoxCeleb2 development set is used to construct the training set, the VoxCeleb1 test set is used to construct the development set, and a portion of the VoxCeleb1 development set is used to construct the test set.

Table~\ref{tab:dataset} presents the statistics of the SSTC training, development, and test sets used for both source and target speakers, detailing the number of speakers and utterances in each set.

\begin{table}[]\centering
\caption{Statistics of source and target speaker datasets in SSTC training, development, and test sets.} 
        \vspace{5pt}
	\setlength{\tabcolsep}{5pt} 
	\renewcommand{\arraystretch}{1.2} 
	\begin{adjustbox}{max width=1\textwidth}
\begin{tabular}{cccc}
\toprule
\textbf{Dataset }                        & \textbf{Subset} & \textbf{\#Speakers} & \textbf{\#Utterances} \\ \midrule
\multirow{3}{*}{\textbf{Source Speaker}} & Train  & 1,172      & 132,553      \\
                                & Dev.    & 40         & 2703         \\
                                & Test   & 40         & 2602         \\ \midrule
\multirow{3}{*}{\textbf{Target Speaker}} & Train  & 5,994       & 1,092,009    \\
                                & Dev.    & 40         & 4,847        \\
                                & Test   & 40         & 4,510        \\ \bottomrule
\end{tabular}
\end{adjustbox}
\label{tab:dataset}
\end{table}

In this study, following the challenge rules, only the train-clean section of Librispeech and the VoxCeleb2 development set, along with the SSTC dataset’s training and development sets, can be used for model training. 
The MUSAN~\cite{snyder2015musan} and RIR Noise~\cite{ko2017study} datasets are used for data augmentation. The detailed usage of these datasets in our proposed method is described in the next section.

\subsection{Setup}
The setup and implementation details of the systems we evaluated in this study and submitted to SSTC are as follows.

\textbf{MFA-Conformer with adaptor}\footnote{\url{https://github.com/SSTC-Challenge/SSTC2024_baseline_system}}: As described in Sections~\ref{sec:2.1} and~\ref{sec:2.2}, the MFA-Conformer~\cite{zhang2022mfa, cai2023leveraging} is used for model training. For the front-end processing, we extract 80 dimensional log Mel-filterbank energies using a frame length of 25ms and a hop size of 10ms. These features are mean normalized before being fed into the deep speaker network.
As mentioned in Section~\ref{sec:2.1}'s back-end part differs slightly from Section~\ref{sec:2.2}. In Speaker Embedding Clustering, an Encoder-Decoder structure is used. In the Encoder, a series of 8 projection layers map the input features to another 176-dimensional hidden representation. This feature representation undergoes Layer Normalization. After pooling the features using attentive statistics pooling, batch normalization is applied again. Finally, a linear layer with an output dimension of 256 maps the features to the final output dimension, and Dropout is applied to prevent overfitting. In adapter-based multi-task Learning, the MFA-conformer model is modified by inserting an adapter after each conformer block within the MFA-Conformer architecture. 

\textbf{ResNet293 with speaker contrastive loss}: The speaker contrastive learning model is trained in three phases. For Phase \Romannum{1}: we follow the standard 293-layer ResNet architecture~\cite{chen2022sjtu} to train the ResNet293 system. The additive angular margin (AAM) loss~\cite{deng2019arcface} is used as the training objective. The scale and margin in the AAM loss are set to 32 and 0.2 respectively. The ResNet293 is trained using the Librispeech train-clean set. In Phase \Romannum{2}, the training set of SSTC 2024 (converted speech) and Librispeech data (source speech) is used to fine-tune the model. Then, the converted speech is used to train the Phase \Romannum{3} model with contrastive loss and AAM loss, with the source speech embedding (extracted by the Phase \Romannum{1} model). The hyper-parameters are set as $K=5$ and $\alpha = 1$.

\begin{table*}[]
	\centering
	\caption{The EER results~(\%) of source speaker verification on development sets.} 
        \vspace{5pt}
	\setlength{\tabcolsep}{4pt} 
	\renewcommand{\arraystretch}{1.2} 
	\begin{adjustbox}{max width=1\textwidth}
		\begin{tabular}{lcccccccccccc}
			\toprule
		\textbf{Method}	& \textbf{Dev-1} & \textbf{Dev-2} & \textbf{Dev-3} & \textbf{Dev-4} & \textbf{Dev-5} & \textbf{Dev-6} & \textbf{Dev-7} & \textbf{Dev-8} & \textbf{Dev-9} & \textbf{Dev-10} & \textbf{Dev-11} & \textbf{Dev-12} \\ \midrule
			\textbf{MFA-Conformer} & 10.781 & 10.191 & 9.034 & 9.232 & 8.699 & 12.984 & 27.137 & 24.061 & 33.360 & 45.674 & 19.734 & 21.668 \\ 
			\hspace{0.5em} \textbf{$+$Adapter}~\cite{li2024database} & 9.535 & 8.971 & 8.290 & 8.013 & 7.861 & 12.919 & 30.129 & 27.999 & 33.189 & 45.060 & 20.599 & 21.564 \\ \midrule
                \textbf{ResNet293} & 9.327 & 9.918 & 7.876 & 7.852 & 7.119 & 10.752 & 24.245 & 21.934 & 29.310 & 40.158 & 19.017 & 20.132 \\ 
                \hspace{0.5em} \textbf{$+\mathcal{L}_{\text{Con}}$} & 8.305 & 9.629 & 7.634 & 7.534 & 6.805 & 10.051 & 23.308 & 20.165 & 27.893 & 39.276 & 18.032 & 18.984 \\ \bottomrule
		\end{tabular}
	\end{adjustbox}
	\label{tab:dev}
\end{table*}

\begin{table}[]\centering
\caption{The EER~(\%) results of source speaker verification on test set.}
\vspace{5pt}
\renewcommand{\tabcolsep}{0.65cm}
\renewcommand\arraystretch{1.25}
\begin{tabular}{lc}
\toprule
 \textbf{Method} & \textbf{EER $\downarrow$} \\\midrule
\textbf{MFA-Conformer} & 20.374   \\
\hspace{0.5em} \textbf{$+$Adapter}~\cite{li2024database} & 20.046   \\\midrule
\textbf{ResNet293} & 18.626   \\
\hspace{0.5em} \textbf{$+\mathcal{L}_{\text{Con}}$} & \textbf{16.788}  \\ \bottomrule
\end{tabular}
\label{tab:test}
\end{table}

\subsection{Evaluation Results}
SLT2024 SSTC evaluates system performance using the Equal Error Rate (EER) metric. 
The trial files for development and test sets consist of three segments: the label (indicating whether the trial is target or non-target), the enrollment utterance ID, and the test utterance ID. For each pair of converted speech utterances in the development and test sets, cosine similarity is computed, and a threshold is applied to determine whether the utterances are from the same source speaker or different source speakers.

Table~\ref{tab:dev} shows the EER results for source speaker verification across 12 different development sets using various models. 
The results indicate that EER varies across VC methods on the development sets because the learned latent source speaker information in the converted speech differs among these different VC methods. 
With the help of the adapter proposed in~\cite{li2024database}, the EER results consistently improve, indicating that this method better adapts to specific datasets. 
Compared to the ResNet293 model, the proposed speaker contrastive learning method significantly improves performance and achieves better EER results than all compared methods. This improvement is due to the enhanced learning of source speaker information through the introduction of speaker contrastive loss, which strengthens source speaker tracing.
 
To further apply in-domain information and supplement data generated by VC methods not seen in the training set, development sets are also added to fine-tune the models. 
The EER results on SSTC test sets 
are shown in Table~\ref{tab:test}, which are consistent with the results on the development sets. Our proposed contrastive learning method achieves an EER of 16.788\%, with a 1.838\% improvement over the ResNet293 model, demonstrating that the latent source speaker information learned through the source speaker contrastive loss is beneficial for source speaker tracing.

\begin{table}[h!]
\centering
\caption{The EER~(\%) results of each competition system on the SSTC test set.}
\vspace{5pt}
\renewcommand{\tabcolsep}{0.85cm}
\renewcommand\arraystretch{1.3}
\begin{tabular}{lc}
\toprule     
\textbf{System}   &  \textbf{EER $\downarrow$} \\ 
\midrule 
\textbf{Proposed (Rank 1st)} &  \textbf{16.788}  \\
Rank 2nd Team & 18.648  \\
Rank 3rd Team &  19.323 \\
Rank 4th Team &  20.027 \\
Rank 5th Team &  20.046 \\ \midrule
Official Baseline &  20.613  \\
\bottomrule    
\end{tabular}
\label{table:leaderboard}
\end{table}
Table~\ref{table:leaderboard} presents the EER results on the SSTC test set for each competition system and the official baseline. Our system achieves a 16.788\% EER on the test set, securing first place in the challenge, which demonstrates the effectiveness of our proposed method in source speaker verification against voice conversion.
Our proposed contrastive learning method yields better results than the official baseline, achieving up to a 3.825\% absolute EER improvement due to the learning of latent source speaker information.

\section{CONCLUSION}
\label{sec:conclu}
In this paper, we propose a source speaker contrastive learning for the source speaker tracing. This method employs speaker contrastive loss to enhance model training, enabling the generation of more discriminative embeddings related to source speakers. The speaker contrastive loss focuses on the latent source speaker information retained in the converted speech. Our approach integrates a speaker embedding extractor trained with converted speech data, learning to capture the source speaker’s characteristics that remain after conversion. By identifying the true source speaker embedding within several distractors of speaker embeddings, the embedding extractor can better learn the potentially possessing source speaker information in the converted speech. 
The proposed system achieves the best performance in the SLT2024 SSTC, with the lowest EER among competition systems.

\bibliographystyle{IEEEbib}
\bibliography{strings,refs}

\begin{thebibliography}{10}

\bibitem{snyder2018x}
David Snyder, Daniel Garcia-Romero, Gregory Sell, Daniel Povey, and Sanjeev Khudanpur,
\newblock ``X-vectors: Robust {DNN} embeddings for speaker recognition,''
\newblock in {\em Proc. ICASSP}. IEEE, 2018, pp. 5329--5333.

\bibitem{desplanques20}
Brecht Desplanques, Jenthe Thienpondt, and Kris Demuynck,
\newblock ``{ECAPA-TDNN: Emphasized Channel Attention, Propagation and Aggregation in TDNN Based Speaker Verification},''
\newblock in {\em Proc. INTERSPEECH}, 2020, pp. 3830--3834.

\bibitem{goodfellow2014explaining}
Ian~J Goodfellow, Jonathon Shlens, and Christian Szegedy,
\newblock ``Explaining and harnessing adversarial examples,''
\newblock {\em Proc. ICLR}, 2014.

\bibitem{carlini2017towards}
Nicholas Carlini and David Wagner,
\newblock ``Towards evaluating the robustness of neural networks,''
\newblock in {\em Proc. SP}. 2017, pp. 39--57, IEEE.

\bibitem{szegedy2013intriguing}
Christian Szegedy, Wojciech Zaremba, Ilya Sutskever, Joan Bruna, Dumitru Erhan, Ian Goodfellow, and Rob Fergus,
\newblock ``Intriguing properties of neural networks,''
\newblock {\em Proc. ICLR}, 2014.

\bibitem{kurakin2016adversarial}
Alexey Kurakin, Ian Goodfellow, and Samy Bengio,
\newblock ``Adversarial examples in the physical world,''
\newblock {\em In Artificial intelligence safety and security}, pp. 99--112, 2018.

\bibitem{wang2019adversarial}
Qing Wang, Pengcheng Guo, Sining Sun, Lei Xie, and John~HL Hansen,
\newblock ``Adversarial regularization for end-to-end robust speaker verification,''
\newblock in {\em Proc. INTERSPEECH}, 2019, pp. 4010--4014.

\bibitem{kreuk2018fooling}
Felix Kreuk, Yossi Adi, Moustapha Cisse, and Joseph Keshet,
\newblock ``Fooling end-to-end speaker verification with adversarial examples,''
\newblock in {\em Proc. ICASSP}. IEEE, 2018, pp. 1962--1966.

\bibitem{wang2020inaudible}
Qing Wang, Pengcheng Guo, and Lei Xie,
\newblock ``Inaudible adversarial perturbations for targeted attack in speaker recognition,''
\newblock {\em Proc. INTERSPEECH}, pp. 4228--4232, 2020.

\bibitem{abdullah2019practical}
Hadi Abdullah, Washington Garcia, Christian Peeters, Patrick Traynor, Kevin~RB Butler, and Joseph Wilson,
\newblock ``Practical hidden voice attacks against speech and speaker recognition systems,''
\newblock {\em Proc. NDSS}, 2019.

\bibitem{wang2023timbre}
Qing Wang, Jixun Yao, Li~Zhang, Pengcheng Guo, and Lei Xie,
\newblock ``Timbre-reserved adversarial attack in speaker identification,''
\newblock {\em IEEE/ACM Transactions on Audio, Speech, and Language Processing}, 2023.

\bibitem{wang2023pseudo}
Qing Wang, Jixun Yao, Ziqian Wang, Pengcheng Guo, and Lei Xie,
\newblock ``Pseudo-siamese network based timbre-reserved black-box adversarial attack in speaker identification,''
\newblock in {\em Proc. INTERSPEECH}, 2023, pp. 3994--3998.

\bibitem{wu2012detecting}
Zhizheng Wu, Eng~Siong Chng, and Haizhou Li,
\newblock ``Detecting converted speech and natural speech for anti-spoofing attack in speaker recognition,''
\newblock in {\em Proc. INTERSPEECH}, 2012, pp. 1700--1703.

\bibitem{wu2015spoofing}
Zhizheng Wu, Nicholas Evans, Tomi Kinnunen, Junichi Yamagishi, Federico Alegre, and Haizhou Li,
\newblock ``Spoofing and countermeasures for speaker verification: a survey,''
\newblock {\em Speech Communication}, vol. 66, pp. 130--153, 2015.

\bibitem{oord2016wavenet}
Aaron van~den Oord, Sander Dieleman, Heiga Zen, Karen Simonyan, Oriol Vinyals, Alex Graves, Nal Kalchbrenner, Andrew Senior, and Koray Kavukcuoglu,
\newblock ``Wavenet: A generative model for raw audio,''
\newblock {\em Proc. ISCA Workshop on SSW 9}, 2016.

\bibitem{ren2019fastspeech}
Yi~Ren, Yangjun Ruan, Xu~Tan, Tao Qin, Sheng Zhao, Zhou Zhao, and Tie-Yan Liu,
\newblock ``Fastspeech: Fast, robust and controllable text to speech,''
\newblock {\em Proc. NeurIPS}, vol. 32, 2019.

\bibitem{cai2023identifying}
Danwei Cai, Zexin Cai, and Ming Li,
\newblock ``Identifying source speakers for voice conversion based spoofing attacks on speaker verification systems,''
\newblock in {\em ICASSP 2023-2023 IEEE International Conference on Acoustics, Speech and Signal Processing (ICASSP)}. IEEE, 2023, pp. 1--5.

\bibitem{li2024database}
Ze~Li, Yuke Lin, Tian Yao, Hongbin Suo, and Ming Li,
\newblock ``The database and benchmark for source speaker verification against voice conversion,''
\newblock {\em arXiv preprint arXiv:2406.04951}, 2024.

\bibitem{zhang2022mfa}
Yang Zhang, Zhiqiang Lv, Haibin Wu, Shanshan Zhang, Pengfei Hu, Zhiyong Wu, Hung-yi Lee, and Helen Meng,
\newblock ``Mfa-conformer: Multi-scale feature aggregation conformer for automatic speaker verification,''
\newblock {\em arXiv preprint arXiv:2203.15249}, 2022.

\bibitem{cai2023leveraging}
Danwei Cai and Ming Li,
\newblock ``Leveraging asr pretrained conformers for speaker verification through transfer learning and knowledge distillation,''
\newblock {\em arXiv preprint arXiv:2309.03019}, 2023.

\bibitem{chen2021again}
Yen-Hao Chen, Da-Yi Wu, Tsung-Han Wu, and Hung-yi Lee,
\newblock ``Again-vc: A one-shot voice conversion using activation guidance and adaptive instance normalization,''
\newblock in {\em Proc. ICASSP}. IEEE, 2021, pp. 5954--5958.

\bibitem{li2023freevc}
Jingyi Li, Weiping Tu, and Li~Xiao,
\newblock ``Freevc: Towards high-quality text-free one-shot voice conversion,''
\newblock in {\em Proc. ICASSP}. IEEE, 2023, pp. 1--5.

\bibitem{gu2021mediumvc}
Yewei Gu, Zhenyu Zhang, Xiaowei Yi, and Xianfeng Zhao,
\newblock ``Mediumvc: Any-to-any voice conversion using synthetic specific-speaker speeches as intermedium features,''
\newblock {\em arXiv preprint arXiv:2110.02500}, 2021.

\bibitem{li2023styletts}
Yinghao~Aaron Li, Cong Han, and Nima Mesgarani,
\newblock ``Styletts-vc: One-shot voice conversion by knowledge transfer from style-based tts models,''
\newblock in {\em Proc. SLT}. IEEE, 2023, pp. 920--927.

\bibitem{park2023triaan}
Hyun~Joon Park, Seok~Woo Yang, Jin~Sob Kim, Wooseok Shin, and Sung~Won Han,
\newblock ``Triaan-vc: Triple adaptive attention normalization for any-to-any voice conversion,''
\newblock in {\em Proc. ICASSP}. IEEE, 2023, pp. 1--5.

\bibitem{wang2021vqmivc}
Disong Wang, Liqun Deng, Yu~Ting Yeung, Xiao Chen, Xunying Liu, and Helen Meng,
\newblock ``Vqmivc: Vector quantization and mutual information-based unsupervised speech representation disentanglement for one-shot voice conversion,''
\newblock {\em arXiv preprint arXiv:2106.10132}, 2021.

\bibitem{zhang2022sig}
Haozhe Zhang, Zexin Cai, Xiaoyi Qin, and Ming Li,
\newblock ``Sig-vc: A speaker information guided zero-shot voice conversion system for both human beings and machines,''
\newblock in {\em Proc. ICASSP}. IEEE, 2022, pp. 6567--65571.

\bibitem{baas2023voice}
Matthew Baas, Benjamin van Niekerk, and Herman Kamper,
\newblock ``Voice conversion with just nearest neighbors,''
\newblock {\em arXiv preprint arXiv:2305.18975}, 2023.

\bibitem{liu2021any}
Songxiang Liu, Yuewen Cao, Disong Wang, Xixin Wu, Xunying Liu, and Helen Meng,
\newblock ``Any-to-many voice conversion with location-relative sequence-to-sequence modeling,''
\newblock {\em IEEE/ACM Transactions on Audio, Speech, and Language Processing}, vol. 29, pp. 1717--1728, 2021.

\bibitem{popov2021grad}
Vadim Popov, Ivan Vovk, Vladimir Gogoryan, Tasnima Sadekova, and Mikhail Kudinov,
\newblock ``Grad-tts: A diffusion probabilistic model for text-to-speech,''
\newblock in {\em International Conference on Machine Learning}. PMLR, 2021, pp. 8599--8608.

\bibitem{lin2021s2vc}
Jheng-hao Lin, Yist~Y Lin, Chung-Ming Chien, and Hung-yi Lee,
\newblock ``S2vc: A framework for any-to-any voice conversion with self-supervised pretrained representations,''
\newblock {\em arXiv preprint arXiv:2104.02901}, 2021.

\bibitem{casanova2022yourtts}
Edresson Casanova, Julian Weber, Christopher~D Shulby, Arnaldo~Candido Junior, Eren G{\"o}lge, and Moacir~A Ponti,
\newblock ``Yourtts: Towards zero-shot multi-speaker tts and zero-shot voice conversion for everyone,''
\newblock in {\em International Conference on Machine Learning}. PMLR, 2022, pp. 2709--2720.

\bibitem{chen2022controlvc}
Meiying Chen and Zhiyao Duan,
\newblock ``Controlvc: Zero-shot voice conversion with time-varying controls on pitch and speed,''
\newblock {\em arXiv preprint arXiv:2209.11866}, 2022.

\bibitem{choi2023diff}
Ha-Yeong Choi, Sang-Hoon Lee, and Seong-Whan Lee,
\newblock ``Diff-hiervc: Diffusion-based hierarchical voice conversion with robust pitch generation and masked prior for zero-shot speaker adaptation,''
\newblock {\em International Speech Communication Association}, pp. 2283--2287, 2023.

\bibitem{kang23b_interspeech}
Wonjune Kang, Mark Hasegawa-Johnson, and Deb Roy,
\newblock ``{End-to-End Zero-Shot Voice Conversion with Location-Variable Convolutions},''
\newblock in {\em Proc. INTERSPEECH 2023}, 2023, pp. 2303--2307.

\bibitem{lim2024wav2vec}
Jaemin Lim and Kiyeon Kim,
\newblock ``Wav2vec-vc: Voice conversion via hidden representations of wav2vec 2.0,''
\newblock in {\em Proc. ICASSP}. IEEE, 2024, pp. 10326--10330.

\bibitem{panayotov2015librispeech}
Vassil Panayotov, Guoguo Chen, Daniel Povey, and Sanjeev Khudanpur,
\newblock ``Librispeech: an asr corpus based on public domain audio books,''
\newblock in {\em Proc. ICASSP}. IEEE, 2015, pp. 5206--5210.

\bibitem{nagrani2017voxceleb}
Arsha Nagrani, Joon~Son Chung, and Andrew Zisserman,
\newblock ``Voxceleb: a large-scale speaker identification dataset,''
\newblock {\em arXiv preprint arXiv:1706.08612}, 2017.

\bibitem{chung2018voxceleb2}
Joon~Son Chung, Arsha Nagrani, and Andrew Zisserman,
\newblock ``Voxceleb2: Deep speaker recognition,''
\newblock {\em arXiv preprint arXiv:1806.05622}, 2018.

\bibitem{snyder2015musan}
David Snyder, Guoguo Chen, and Daniel Povey,
\newblock ``Musan: A music, speech, and noise corpus,''
\newblock {\em arXiv preprint arXiv:1510.08484}, 2015.

\bibitem{ko2017study}
Tom Ko, Vijayaditya Peddinti, Daniel Povey, Michael~L Seltzer, and Sanjeev Khudanpur,
\newblock ``A study on data augmentation of reverberant speech for robust speech recognition,''
\newblock in {\em Proc. ICASSP}. IEEE, 2017, pp. 5220--5224.

\bibitem{chen2022sjtu}
Zhengyang Chen, Bei Liu, Bing Han, Leying Zhang, and Yanmin Qian,
\newblock ``The sjtu x-lance lab system for cnsrc 2022,''
\newblock {\em arXiv preprint arXiv:2206.11699}, 2022.

\bibitem{deng2019arcface}
Jiankang Deng, Jia Guo, Niannan Xue, and Stefanos Zafeiriou,
\newblock ``Arcface: Additive angular margin loss for deep face recognition,''
\newblock in {\em Proceedings of the IEEE/CVF conference on computer vision and pattern recognition}, 2019, pp. 4690--4699.

\end{thebibliography}

\end{document}